# Suppressing Mechanical Property Variability in Recycled Plastics *via* Bio-inspired Design

Dimitrios Georgiou[1], Danqi Sun[1], Xing Liu[2], Christos E Athanasiou[1,*]

1. Daniel Guggenheim School of Aerospace Engineering, Georgia Institute of Technology, Atlanta, GA 30332, USA

2. Department of Mechanical and Industrial Engineering, New Jersey Institute of Technology, Newark, NJ 07102, USA

*Corresponding author: Christos E Athanasiou (athanasiou@gatech.edu)

## Abstract

The escalating plastic waste crisis demands global action, yet mechanical recycling – currently the most prevalent strategy – remains severely underutilized. Only a small fraction of the total plastic waste is recycled in this manner, largely due to the significant variability in recycled plastics' mechanical properties. This variability stems from compositional fluctuations and impurities introduced throughout the materials' lifecycle and the recycling process, deterring industries with stringent product specifications from adopting recycled plastics on a wider scale. To overcome this challenge, we propose a composite structure inspired by nacre's microstructure – a natural material known for its exceptional mechanical performance despite its inherent randomness across multiple length scales. This bio-inspired design features stiff recycled plastic platelets ("bricks") within a soft polymeric matrix ("mortar"). We use a tension-shear-chain model to capture the deformation mechanism of the structure, and demonstrate, through a case study of commercial stretch wrap, that the proposed design reduces variability in effective elastic modulus by 89.5% and in elongation at break by 42%, while achieving the same modulus as the virgin stretch wrap material. These findings highlight the potential of the proposed bio-inspired design to enhance the mechanical performance of recycled plastics, but also demonstrate that a universally applicable, chemistry-agnostic approach can substantially broaden their applications, paving the way for sustainable plastic waste management.

## 1 Introduction

500 million tons of plastics are produced annually, with this amount projected to reach 1 billion tons per year by 2050 [1,2]. Although recycled polymers (recyclates) are necessary to transition away from fossil-based products, currently less than 9% of the generated plastic waste is being recycled (even though 90% is recyclable), with the remaining 91% accumulating in landfills, dumps, and the natural environment.

Mechanical recycling, *i.e.,* shredding, melting, and reforming plastic waste, is currently the most resource efficient plastics recycling option [3]. However, mechanically recycled plastics often exhibit degraded and inconsistent mechanical performance compared to their virgin counterparts. This variability in mechanical properties arises even when using identical polymer feedstocks [4], limiting their applicability in many industries and ultimately leading to materials that can no longer be recycled [5]. These challenges primarily stem from compositional fluctuations and impurities introduced during the recycling process, as well as from exposure to environmental factors and degradation of material properties throughout the material lifecycle. For example, La Mantia and Vinci observed the erratic fracture behavior of post-consumer polyethylene terephthalate (PET) bottles under dry and humid conditions across multiple recycling cycles [6]. They reported a reduction in elongation at break from 36% to 20% and an increase in Young's modulus from 1300 MPa to 1580 MPa for humid bottles over five recycling cycles. Eriksen et al. observed significant variability in the strength and tensile strain of recycled high-density polyethylene (rHDPE) and recycled low-density polyethylene (rLDPE) sourced from household waste, ranging from 22 to 50 MPa and 11 to 18%, respectively, attributing it to the heterogeneity of the waste stream [7]. Cecon et al. evaluated the performance of mixed plastic waste sourced from U.S. material recovery facilities (MRFs) and observed significant variability in the Young's modulus of recyclates across different MRFs, with the highest variability reported for rHDPE, ranging from 380 MPa to 1250 MPa [8]. This variability was linked to differences in bale composition and the inconsistent quality of input streams. Overall, there is consensus in the literature that heterogeneities and compositional fluctuation are inherent features of recyclates.

On the contrary, natural materials, such as nacre and bamboo, do not exhibit significant variability in their mechanical performance and properties despite the compositional fluctuations and potential impurities introduced during their lifecycle [9–12]. This is due to their brick-and-mortar architecture, which combines stiff mineral platelets with thin proteinic layers, exhibiting remarkable insensitivity to flaws at the nanoscale [13]. Due to this architecture, these materials feature energy dissipation mechanisms, such as crack deflection, interlocking tiles, and crack bridging, effectively preventing catastrophic failure, while the proteinic layers alone accommodate plastic deformation and redistribute stress to ensure mechanical robustness despite microscopic defects and microstructural randomness [14–19].

Inspired by nacre's hierarchical structure, we propose a composite structure design that aims to suppress variability in recyclates. In this bio-inspired design, recyclates serve as stiff platelets (bricks), while soft polymeric interfaces (mortar) accommodate the majority of deformation. When the composite structure is subjected to loads, the soft interfaces effectively transfer stress between the stiff platelets, ensuring uniform stress distribution throughout the structure. By identifying the key mechanical properties of the platelets and quantifying their variability based on extensive literature review (Figure 1A), we employ Monte Carlo simulations within a tension-shear-chain model to predict the effective elastic modulus and elongation at break of the structure. We investigate the influence of geometric parameters and material properties on the overall performance and variability. Finally, we compare the initial variability of the recyclate platelets to that of the final brick-mortar structure, demonstrating a reduction of up to 89.5% in variability while achieving the same modulus as that of the virgin platelet material and showcasing the design's applicability for industrial stretch wrap. An overview of our approach is presented in Figure 1.

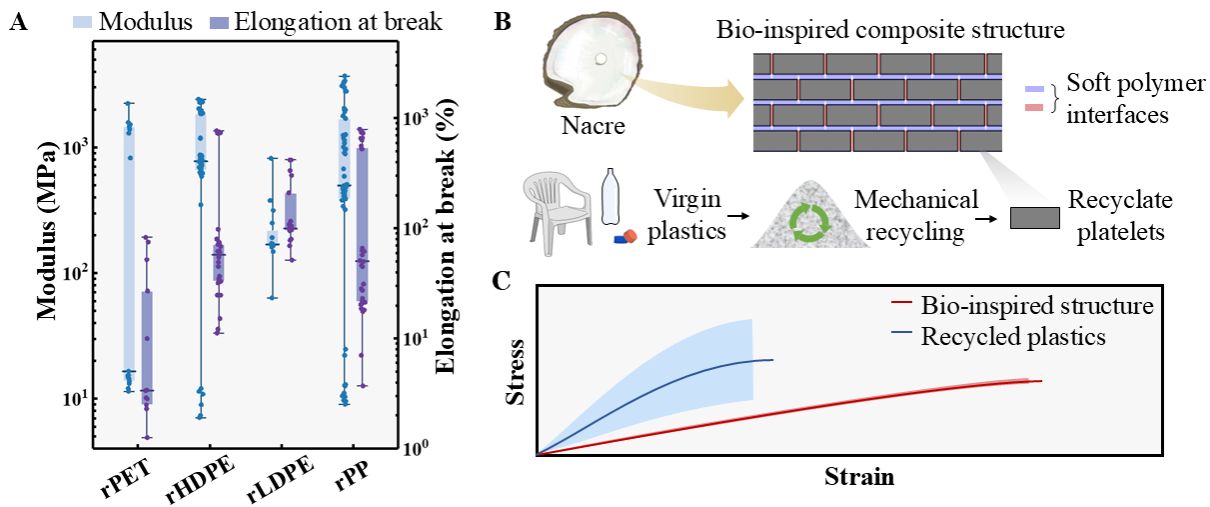

**Figure 1. Suppressing variability through a bio-inspired composite structure. (A)** Summary of literature data [6–8,20–27] demonstrating the significant variability in the mechanical properties of recyclable plastics (Details in Figure S1). Plot of the modulus and the elongation at break of rPET, rHDPE, rLDPE and rPP. **(B)** Schematic of a nacre-inspired recyclable composite structure. Blocks of recycled plastics with uncertainty in their mechanical behavior are represented as hard bricks and are 'glued' by a softer polymer. **(C)** Conceptual schematic illustrating the broader uncertainty in the mechanical properties of recyclates compared to the narrower uncertainty that is desired in industrial applications. Solid lines and shadow bands represent mean values and standard deviation, respectively.

## 2  Modelling of bio-inspired recyclate structure

### 2.1  Tension-Shear-Chain network model

We employ a discrete element approach, similar to Ref. [18], that incorporates three types of elements: Cohesive Tensile (CT), Cohesive Shear (CS), and Hard elements. The model creates a staggered network of platelets with equal overlap, as illustrated in Figure 1B. The network is characterized by the number of platelets in each layer, $N$, and the number of layers, $M$. An $\{N, M\}$ network comprises $N \times M$ hard elements.

CT elements connect hard elements within the same layer (horizontal direction), accounting for deformation across the same layer while CS elements link platelets between different layers (vertical or diagonal direction), capturing shear between the staggered layers of the network. The total number of CT and CS elements in an $\{N, M\}$ network is $N_{CT} = (N-1)M$ and $N_{CS} = (2N-1)(M-1)$, respectively. The CT and CS interfaces are respectively characterized by a bilinear cohesive constitutive law defined by three parameters: the maximum traction, $\sigma_{max,CT}$ and $\tau_{max,CS}$, which represent the peak stresses the interface can sustain before damage initiation; the elastic separation, $\delta_{e,CT}$ and $\delta_{e,CS}$, where traction reaches its maximum during elastic loading and the interface is in the onset of damage; and the critical separation, $\delta_{cr,CT}$ and $\delta_{cr,CS}$, signaling full decohesion of the interface. Hard elements are assumed to be linear elastic, characterized by a single material parameter, the Young's modulus, $E_{hard}$, while their contraction in the vertical direction, during uniaxial loading in the longitudinal direction, is considered negligible.

The network is modeled as a system of two-node elements, with each element resembling a generally nonlinear spring. Only horizontal displacements are considered under uniaxial tension. The system features $N_{DOF} = 2(N \times M)$ degrees of freedom (DoFs), corresponding to two horizontal displacement components per hard element (Figure 2A). From the nodal displacements, the displacement jumps across the elements are calculated as, $\delta = u_i - u_j \; \forall \; i \neq j \; \epsilon \; [1, N_{DOF}]$. The stress state and damage condition of each element are determined from the displacement jumps using a one-to-one relationship between deformation and the corresponding stress, as defined by Eqs. (1-4). For the hard elements, 1D line elements are employed (Eq. 1 and Figure 2B), and their performance is compared, across various displacement field states, to the stress predicted by the shear-lag model (Eq. 2) [28], where the contributions of neighboring CT and CS elements determine the average stress in a platelet. The results demonstrate accurate predictions for elements in the inner layers of the structure, while the shear-lag model estimates slightly higher stresses in the top and bottom layers. The stresses in the shear and tensile interfacial elements are determined directly from the traction-separation law, based on whether the element is intact, damaged, or fully destroyed (Eqs. 3,4):

$$\sigma_{Hard} = \frac{E_{hard} A}{l} \delta \qquad (1)$$

$$\sigma_{lag} = \sigma_{CT} + \frac{l}{h} \tau_{CS} \qquad (2)$$

$$\sigma_{CT} = \begin{cases} K_{CT}^+ \delta, & \delta \leq \delta_{e,CT} \\ K_{CT}^- \delta, & \delta_{e,CT} < \delta < \delta_{cr,CT} \\ 0, & \delta \geq \delta_{cr,CT} \end{cases} \qquad (3)$$

$$\tau_{CS} = \begin{cases} K_{CS}^+ \delta, & \delta \leq \delta_{e,CS} \\ K_{CS}^- \delta, & \delta_{e,CS} < \delta < \delta_{cr,CS} \\ 0, & \delta \geq \delta_{cr,CS} \end{cases} \quad (4)$$

where $l$ is the platelet length and $h$ is the platelet height, and $A$ the cross sectional area of the platelet. For the interfaces, we employ a variable stiffness approach that captures the bilinear nature of the cohesive law (Figures 2C, D). The stiffness of an element during elastic deformation, denoted as $K^+$, is defined as the ratio of the maximum traction, $\sigma_{max,CT}, \tau_{max,CS}$, to the elastic deformation limit, $\delta_{e,CT}, \delta_{e,CS}$, for both CT and CS elements respectively (Eqs. 5,6). To model damage, we assume that the stiffness of the interfacial elements decreases as damage progresses; therefore, the element showcases softening. The reduced stiffness, $K^-$, of each element depends on the current displacement jump across the element and is calculated as the slope of the line connecting the origin to the point $[s, \delta]$ on the traction-separation curve (Eqs. 7,8 and Figures 2C, D).

$$K_{CT}^+ = \frac{\sigma_{CT}}{\delta_{e,CT}} \quad (5)$$

$$K_{CS}^+ = \frac{\tau_{CS}}{\delta_{e,CS}} \quad (6)$$

$$K_{CT}^- = \frac{\delta_{cr,CT}}{\delta} \frac{\delta_{cr,CT} - \delta}{\delta_{cr,CT} - \delta_{e,CT}} K_{CT}^+ \quad (7)$$

$$K_{CS}^- = \frac{\delta_{cr,CS}}{\delta} \frac{\delta_{cr,CS} - \delta}{\delta_{cr,CS} - \delta_{e,CS}} K_{CS}^+ \quad (8)$$

During loading, when strain localization occurs at certain interfaces, other elements can experience unloading. For damaged elements, the stress corresponding to the displacement jump during unloading deviates from the bilinear traction-separation law, being instead determined by the reduced stiffness defined in the second branch of the piecewise relation in Eqs. (3,4). The value of the reduced stiffness, $K^-$, is predefined based on the displacement jump, $\delta$, prior to unloading (dashed line in Figures 2C, D).

To solve the system, we employ an iterative Newton scheme to enforce force equilibrium at each node. The equilibrium condition at each node is realized by iteratively updating the nodal displacements until the residual forces converge within a predefined tolerance. This approach is feasible due to the uniaxial tensile loading of the model and the underlying assumptions, which enable a streamlined mapping of stresses to forces while considering only horizontal components in the equilibrium equations. The nodal displacements at each iteration are updated according to Eq. (9):

$$\boldsymbol{u}_{new} = \boldsymbol{u}_{old} - [\nabla \boldsymbol{R}]^{-1} \boldsymbol{R} \quad (9)$$

where $R$ is the residual force vector, and $\nabla R$ is the Jacobian matrix. The iterations continue until the second norm of the residual vector, $||R||_2$, is less than $\varepsilon_R = 10^{-6}$, ensuring convergence. The residual vector, $R$, represents the imbalance of forces at each node and is defined in Eq. (10), where $F_{ext}$ is the vector of externally applied forces, and $F_{int}$ is the vector of internal forces arising from the element stresses.

$$R = F_{ext} - F_{int} \qquad (10)$$

The internal forces are computed by summing the contributions from all connected elements at each node, based on the stresses and orientations of the elements. For the uniaxial tension setup, $R$ contains only horizontal force components. The internal forces are derived from the stress states of the CT, CS, and hard elements. These stresses are calculated based on the current displacement jumps using the traction-separation laws for the CT and CS elements and the linear constitutive model for the hard elements (Eqs. 1,3,4 and Figure 2B-D). The Jacobian matrix, $\nabla R$, accounts for the nonlinear relationship between the forces and the displacements, incorporating the stiffness variation of damaged elements.

Boundary conditions are imposed by fixing the horizontal displacements of nodes along the left boundary and applying a specified displacement load, $\Delta$, to nodes on the right boundary (Figure 2A). The displacement load is incrementally increased in small steps until the predefined total applied elongation $\Delta$ is reached. This incremental loading ensures stability during the simulation and allows the system to accurately capture the nonlinear deformation and damage evolution in the network. The algorithm outputs the displacement values for all DoFs at each displacement increment, along with the corresponding element stresses, damage states (intact, damaged, unloaded, or destroyed), and the reaction force probed at the right boundary. These values are utilized to compute the structure's effective modulus, elongation at break, and other relevant metrics, such as fracture stress.

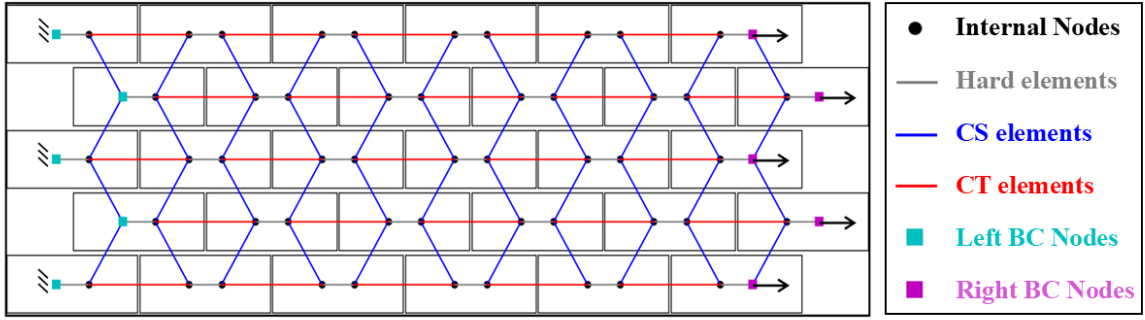

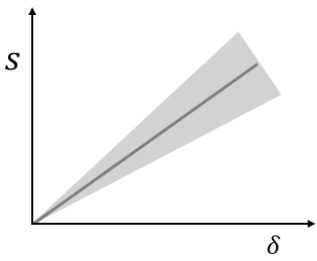
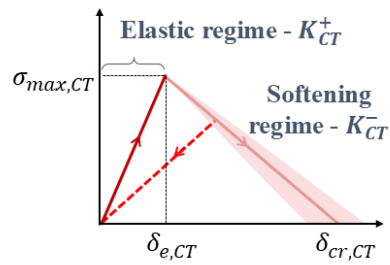
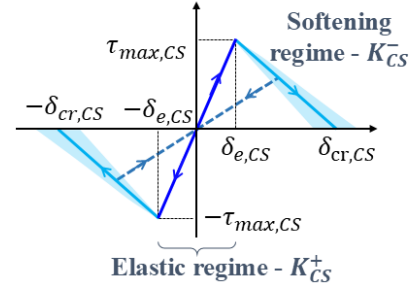

**Figure 2. Components of the Discrete Network. (A)** Connectivity of a {5,5} network, featuring Hard Elements (grey), CT elements (red), and CS elements (blue). Boundary nodes are highlighted: teal for fixed nodes and purple for nodes with externally applied displacement. **(B)** Stress-strain curve for uniaxial tension of the Hard Elements. **(C)** Constitutive cohesive law of the CT elements. **(D)** Constitutive cohesive law of the CS elements. Solid lines and shaded bands represent the mean values and standard deviations, respectively, highlighting the variability in the properties of the recyclates (Section 2.2), while dashed lines indicate the unloading path of the material.

## 2.2 Platelet Mechanical Property Variability

We introduce variability in the platelets by assigning stochastic mechanical properties to the hard elements, reflecting the inherent randomness of recycled plastics, by drawing inspiration from the stochastic approaches in the models proposed by Luo and Bažant [29] and Yan et al [18]. We sample the Young's modulus $E_{hard}$ using a two-parameter Gaussian probability distribution (Eq. 11 and Figure 2B), which is parameterized based on experimental data from the literature [6–8,20–27] while we model the properties of the elastic regime of interfaces to be deterministic. The probability density function (PDF) of the Gaussian distribution is given by Eq. (11):

$$F(x) = \frac{1}{\sqrt{2\pi\sigma^2}} e^{-\frac{(x-\mu)^2}{2\sigma^2}} \tag{11}$$

Where $x$ is the random variable, and $\mu, \sigma$ is the mean value and the standard deviation of the population respectively. To compute the ensemble effective modulus $E_{eff}$ of the network, we evaluate the effective modulus and elongation at break for multiple realizations of the network, each generated with randomized mechanical properties for each of the platelets. For every realization, the effective modulus is computed as the slope of the linear region of the stress-strain curve (Eq. 12).

$$E_{eff} = \frac{\sigma_{dmg}}{\Delta} L \tag{12}$$

Here, $\sigma_{dmg}$ represents the stress at the last increment before damage is detected in the structure. The initial length of the structure, $L$, corresponds to the cumulative length of the platelets and CT interfaces along a single layer while considering the staggered arrangement of layers. The $E_{eff}$ values are then aggregated across all realizations to compute the ensemble statistics: the mean effective modulus, $\mathbb{E}[E_{eff}]$, is calculated as the average of the individual $E_{eff}$ values, while the variance of the $E_{eff}$ values, computed as $\mathbb{V}[E_{eff}] = \sigma^2$, indicates the spread around the mean.

We further consider that the inherent variability of the recyclates introduces uncertainty in the softening regime of the cohesive elements (Shadow Figure 2C, D), as the interfacial behavior is closely tied to the properties of the substrate [30,31]. We sample the critical separation parameters of the interfaces, $\delta_{cr,CT}$ and $\delta_{cr,CS}$ based on Eq. (11), and evaluate the elongation at break $\varepsilon_{max}$ for multiple realizations of the network. The elongation at break is defined as the ratio of the change in the structure's length to its original length at the point of failure. The $\varepsilon_{max}$ values obtained are used to compute the ensemble statistics, following the same approach as with $E_{eff}$. We quantify the stochasticity of the network's properties by the ratio of the Coefficient of Variation ($CV$, Eq. 13) of the material to that of the ensemble property ($E_{eff}, \varepsilon_{max}$).

$$CV = \frac{\sigma}{\mu} \tag{13}$$

## 2.3 Model Validation through simple cases

To validate the proposed model, we analyze its behavior under two simple scenarios: the *serial chain* and *alternating column* configurations focusing on $E_{eff}$. In the serial chain case, the system consists solely of CT interfaces and hard elements, all connected in series within a single layer. The overall stiffness decreases as the applied force from the displacement load is transmitted through all elements, with the most compliant constituents undergoing the largest deformation and dominating the structural response. In the alternating column case, the network's connectivity consists of a combination of parallel and series configurations, resulting from the alternating arrangement of platelets. The stiffness is determined by the combined contribution of the deterministic CS interfaces and stochastic hard elements, with the compliant CS interfaces absorbing most of the deformation. In both cases, the structure's stiffness is heavily influenced by the soft interfaces.

To obtain an analytical expression of the overall stiffness of the structure for each of the two cases we analyze the connectivity of the elements. In the serial chain case, the system can be approximated as a collection of springs in series. Assuming that $K_{HARD_i}$ is the stiffness of the i-th hard platelet, $K_{CT}^+$ the stiffness of any CT element, the overall stiffness is described by Eq. (14)

$$\frac{1}{K_{total}} = \sum_{i=1}^{N} \frac{1}{\overline{K_{HARD_i}}} + (N-1)\frac{1}{\overline{K_{CT}^+}} \tag{14}$$

where $w$ is the thickness of the structure, $h$ is the height of the platelets and $\overline{K_{HARD_i}} = E_{HARD_i}\left(\frac{wh}{l}\right)$, $\overline{K_{CT}^+} = K_{CT}^+(wh)$ are measures of stiffness with consistent units.

An analytical expression for the overall stiffness of the structure in the alternating column case is less straightforward. The simplest configuration, consisting of only three layers, can be analyzed using only a combination of series and parallel spring connections. In general, for structures with an odd number of layers, $M$, an analytical solution can still be determined by extending this approach. This is possible by using the compliance $C = 1/K$ of the elements instead of the stiffness directly and considering transformations between Delta ($\Delta$) and Wye ($Y$) spring connections in the network (Figure S2). Eventually the original network of size $\{N, M\}$ decomposes to an equivalent network of series and parallel connections. For a network consisting of *five* layers the overall stiffness is given by Eq. (15)

$$K_{total} = \frac{6\overline{C_{HARD_i}} + 4\overline{C_{CS}}}{5\overline{C_{HARD_i}}^2 + 5\overline{C_{HARD_i}}\,\overline{C_{CS}} + \overline{C_{CS}}^2} \tag{15}$$

Where $\overline{C_{HARD_i}} = 1/\overline{K_{HARD_i}}$ $\overline{C_{CS}} = 1/\overline{K_{CS}^+}$, are the compliances of the i-th hard element and any CS interface respectively. Here, $\overline{K_{CS}^+} = K_{CS}^+\left(\frac{wl}{2}\right)$ is a measure of stiffness with consistent units and $l$ is the length of each platelet.

We validate our model by comparing the results obtained from Eqs. (14,15) to the corresponding results of the model simulations. We select rHDPE as a model material due to the availability of more data compared to other recycled plastics. Assuming that $E_{hard}$ follows a normal distribution, we calculate its mean value from the literature [6,8,20–27] as $\mathbb{E}[E_{hard}] = 1213.84$. The standard deviation of these samples is $\sigma = 654$ (Sample size = 40). To ensure $E_{eff}$ corresponds to a reasonable normal distribution, we set $\sigma = 200$, representing an elevated variability level featuring a $CV$ of 16.48%. Therefore, in the validation cases $E_{hard}$ is sampled using Eq. (11) with parameters $\mathcal{N}(1213.84, 200^2)$.

Deriving an analytical expression for the distribution of $E_{eff}$ is challenging as the normal distribution is not always reversible, consequently, to quantify the variability in the structure's effective modulus, we employ Monte Carlo simulations. A total of $10^6$ simulations are conducted for a network of size {5,1} in the serial chain configuration and {1,5} in the alternating column configuration to obtain the distribution of $K_{total}$ (Eqs. 14, 15). The effective modulus, $E_{eff}$, is obtained through Eq. (16).

$$\sigma = \underbrace{K_{total}\left(\frac{L}{wh}\right)}_{E_{eff}} \frac{\Delta}{L} \tag{16}$$

We then perform $10^6$ simulations using the developed computational model to monitor the stiffness of the structure up to the point damage is detected (Eq. 12). The model's parameters used in the validation studies are presented in Table 1. The interfacial cohesive properties are assigned values representative of soft materials [32,33], with the interfacial thickness set to 0.25 mm and 0.5 mm for CT and CS interfaces, respectively. These thicknesses are chosen to align with manufacturing constraints while ensuring compatibility with the selected platelet dimensions. The platelet length, height, and width are designed to create a tensile specimen with overall dimensions of $\{L, H, W\} = \{152.5, 32.5, 10\}$ corresponding to a network size of $\{N, M\} = \{10, 5\}$.

Table 1. Model Parameters for the validation cases

| Parameter | Value |
|---|---|
| CT elements $\{\sigma_{max,CT}, \delta_{e,CT}, \delta_{cr,CT}\}$ (MPa/mm/mm) | {4, 0.2, 0.5} |
| CS elements $\{\tau_{max,CS}, \delta_{e,CS}, \delta_{cr,CS}\}$ (MPa/mm/mm) | {4, 0.2, 0.5} |
| Platelet Length $l$ (mm) | 15 |
| Platelet Height $h$ (mm) | 6 |
| Platelet Width $w$ (mm) | 10 |

The obtained distributions closely align, with their mean value being consistent to the second decimal point, and the standard deviation differing by less than 4% between the two cases (Figure 3). The distributions exhibit a slight left-tailed inclination, attributed to the greater influence of softer platelets to the structure's stiffness (Figure S3 and Table S1). This preliminary validation of the model also demonstrates the composite structure's capability to significantly reduce randomness, as the variability, quantified by the $CV$, has been reduced by over 70% in both cases

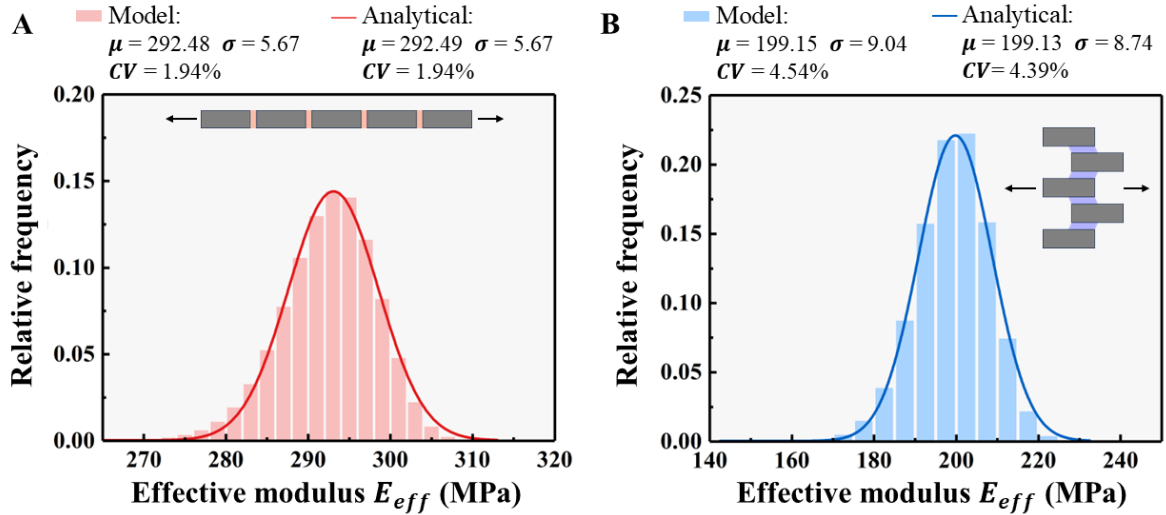

**Figure 3. Validation of the model through limit cases**. The cumulative probability distribution functions for the analytical solution are compared with the results of Monte Carlo simulations from the proposed model. **(A)** *Serial Chain Case*: The stiffness is primarily governed by the most compliant element, with the distribution showing excellent agreement between analytical predictions and model simulations up to the second decimal digit. The structure demonstrates 88% less variability with a $CV$ ratio of $CV_{serial}/CV_{rHDPE} = 0.1177$. **(B)** *Alternating Column Case*: The stiffness results from a combination of parallel and series connections, with both analytical and simulation results demonstrating consistent mean and variability. The structure demonstrates 73% less variability with a $CV$ ratio of $CV_{AC}/CV_{rHDPE} = 0.2664$.

## 3 Results

With the model validated, we proceed to investigate the influence of key parameters on the network's effective modulus. Specifically, we examine how network size, platelet dimensions, and interfacial stiffness impact the structure's overall elastic response, as these factors directly affect stress distribution, strain localization mechanisms, and failure. Additionally, we employ those findings to demonstrate the potential application of our bio-inspired composite in industrial stretch wrap where stretchability without variability and with minimal damage is required.

## 3.1 Effect of model's parameters on the structure's effective modulus

To investigate the effect of network size on $E_{eff}$, we conduct a sensitivity analysis of the parameters expected to have the greatest influence on the system's modulus: the number of platelets per layer $N$, the number of layers $M$, the geometric properties of the platelets $(l, h)$ and the interfacial stiffness $K^+$. The arrangement of the platelets is considered uniform and periodic. We divide these parameters into three categories: the network size (Figure 4A, B), the platelet dimensions (Figure 4C) and the interfacial properties (Figure 4D). To efficiently perform a high number of simulations, we conduct a convergence analysis to determine the sample size required for achieving reliable results for the model's variability. Based on this analysis, we select 200 simulations as the optimal sample size (Figure S4).

First, we vary the network size, defined by the number of platelets per layer, $N$, and the number of layers, $M$, while keeping all other parameters constant. The ensemble $E_{eff}$ is computed for various network sizes $\{N, M\}$ to evaluate their impact on the structural properties while the remaining model parameters are listed in Table 1. Smaller networks exhibit higher variability, which is expected given the reduced number of interfaces relative to platelets (Figure 4A). As the network size increases, the deterministic interfaces begin to dominate, effectively eliminating the stochasticity in the properties of the platelets (Figure 4B). Additionally, increasing the number of platelets reduces the overall stiffness for a given number of layers, as the added platelets resemble the serial chain configuration, where the equivalent stiffness is lower than the stiffness of individual constituents. In contrast, increasing the number of layers for a given number of platelets has minimal effect on stiffness due to the more complex connectivity of the elements. This behavior agrees with Eq. (15), where element contributions appear in both the numerator and denominator of the total stiffness.

Second, we examine the sensitivity of the network's properties to the geometric parameters of the platelets (Figure 4C). We vary the platelet height, $h$, and length, $l$, while maintaining the interfacial properties listed in Table 1. The $\{10,5\}$ network size is chosen for its balance between reduced variability and adequate stiffness. Increasing the platelet length leads to higher stiffness, as the contribution of CS elements is influenced by the aspect ratio of the platelets [28]. However, while increasing the platelet height enhances the stiffness of the CT interfaces, it simultaneously increases the cross-sectional area, minimizing its overall contribution to the structure's stiffness.

Finally, we examine the effect of interfacial stiffness (Figure 4D) by varying interfacial strength while keeping the elastic separation limit parameter constant. The properties in Table 1 and the $\{10,5\}$ network size are maintained for consistency. The results show that the CT interfaces have the greatest impact on the structure's stiffness, as they directly resist deformation during loading. In contrast, the CS interfaces play a secondary role due to their connectivity. As CT stiffness increases, the overall stiffness of the structure becomes dominated by the tensile behavior of the CT interfaces, with the CS interfaces primarily responsible for stress redistribution rather than contributing significantly to the effective modulus.

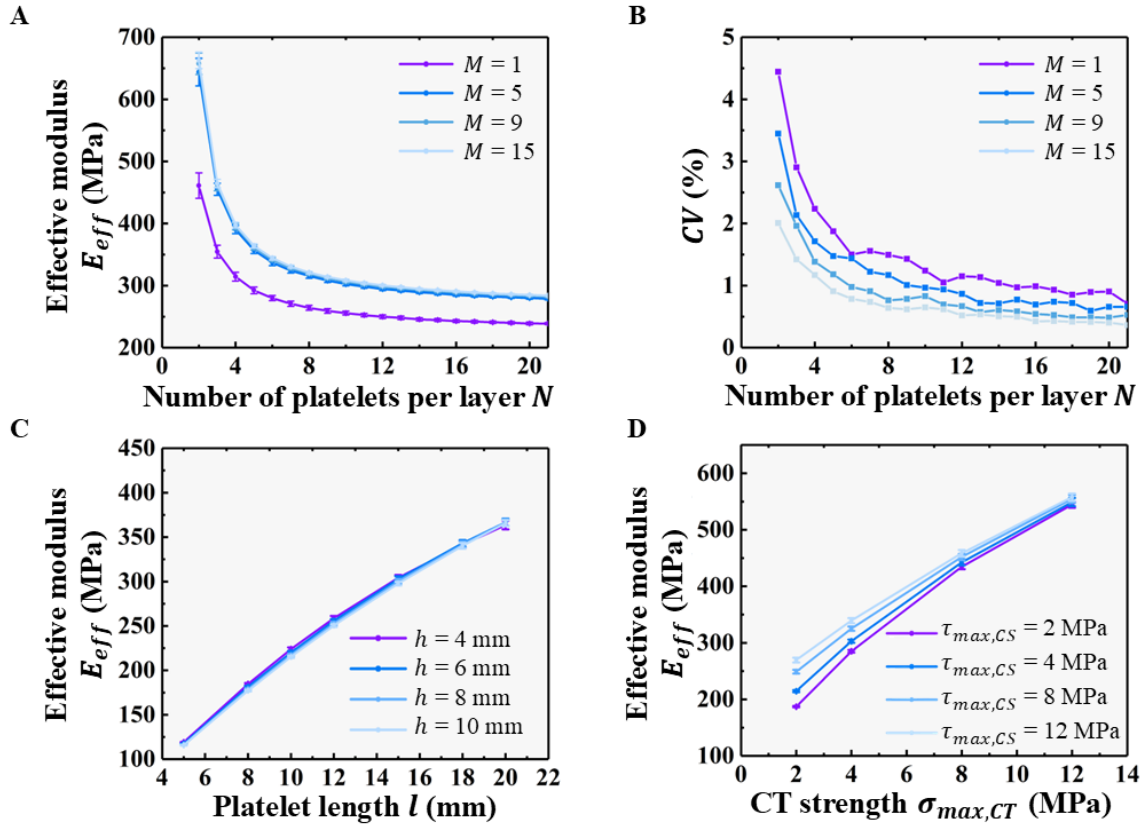

**Figure 4. Sensitivity Analysis of the network's parameters (A)** *Network Size*: Variation in the effective modulus, $E_{eff}$, with the number of platelets per layer $N$ and the number of layers $M$. The results highlight the convergence of structural properties for larger network sizes **(B)** *Variability Reduction*: The $CV$ decreases with an increasing network size. **(C)** *Platelet dimensions*: Various combinations of platelet height, $h$, and length, $l$, are analyzed, highlighting the impact of platelet length on the structure's properties. **(D)** *Interfacial Properties*: Combinations of interfacial stiffness illustrate the relative contributions of CT and CS elements to the overall stiffness with the CT elements dominating the structure's $E_{eff}$.

### 3.2  Case Study: Recycling of Industrial Stretch Wrap

Industrial stretch wrap is a thin, stretchable film, predominantly made from low-density polyethylene (LDPE), widely used to secure goods during packaging and shipping, helping to prevent damage and tampering. The annual production of industrial stretch wrap in the United States is 4,000,000 tons [34], with a market demand valued at 4.47 billion USD [35]. While stretch wrap is recyclable, a barrier to its reuse lies in the significant variability in the mechanical properties of recycled LDPE. This variability, driven by both material degradation during use and further inconsistencies introduced in the recycling process, undermines the reliability required for industrial applications.

During its lifecycle, environmental stressors like prolonged UV exposure degrade LDPE's mechanical properties [36,37], causing its Young's modulus to decrease by 1.6% annually under solar radiation [38]. Recycling intensifies this problem leading to further property degradation [39,40]. Although rLDPE can exhibit increased stiffness, its tensile strength and elongation at break are often compromised. Furthermore, even under controlled recycling conditions, the mechanical properties of rLDPE vary significantly depending on the recycling content or number of recycling cycles [41–43], making it difficult to achieve the consistency needed for industrial stretch wrap applications. For example, in a low contamination recycling scenario, rLDPE is estimated to have a Young's modulus of 234.2 MPa (119% of its pre-recycling value) [42], but in highly contaminated cases, this can even reach 419.2 MPa (213% of its pre-recycling value), with both instances creating unpredictable performance [43]. This variability extends to the material's stretchability before the onset of damage, and the performance of recycled LDPE deviates from that of the virgin material, making it unsuitable for the requirements of industrial stretch wrap.

To meet the performance requirements for industrial applications, rLDPE must exhibit properties comparable to those of virgin LDPE, including an effective Young's modulus of approximately 200 MPa [44,45], a working strain of at least 0.1 to ensure sufficient stretchability, and low variability for predictable performance. To address this need, we suggest employing rLDPE as hard platelets to fabricate a bio-inspired rLDPE composite structure of size {8,5} with soft polymers as interfacial adhesives. The interfacial thickness is set to 1 mm for both interfaces while model parameters for this design are selected to minimize the variability of the structure's properties while maintaining desirable mechanical response based on the results of Section 3.1 (Figure 5A and Table 2). The mechanical properties of rLDPE are assumed to follow a normal distribution. Consequently, the critical separation displacement of the interfaces is assumed to be also normally distributed with a $CV$ of 15%. We sample $E_{hard}$ as $\mathcal{N}(326.7, 30.8^2)$. The yield strain of rLDPE ranges from 0.017 to 0.138.

**Table 2. Model Parameters for the case study**

| Parameter | Value |
|---|---|
| CT elements $\{\sigma_{max,CT}, \delta_{e,CT}, \delta_{cr,CT}\}$ (MPa/mm/mm) | $\{4, 0.15, 1.2 \pm 0.54\ (3\sigma)\}$ |
| CS elements $\{\tau_{max,CS}, \delta_{e,CS}, \delta_{cr,CS}\}$ (MPa/mm/mm) | $\{6, 0.6, 1.5 \pm 0.675\ (3\sigma)\}$ |
| Platelet Length $l$ (mm) | 18 |
| Platelet Height $h$ (mm) | 6 |
| Platelet Width $w$ (mm) | 2.5 |

We define the working strain $\varepsilon_w$ of the bio-inspired rLDPE composite structure as the strain at its fracture. We conduct 200 simulations to obtain the ensemble $E_{eff}$, $\varepsilon_w$ and their variability. The effective modulus and working strain of the bio-inspired structure both meet the requirements of the virgin LDPE. The obtained effective modulus of the bio-inspired structure is 202.1±2.0 MPa, and the working strain is 0.102±0.009. The bio-inspired structure demonstrates 89.5% less variability in effective modulus (Figure 5B). As shown in Figure S1B, the yield strain of rLDPE reported in the literature is 0.017, 0.03, and 0.138, respectively. Beyond the yield strain, rLDPE undergoes significant plastic deformation, making it unsuitable for reuse. Therefore, we consider the yield strain of rLDPE as its working strain $\varepsilon_w$. The working strain of the bio-inspired rLDPE composite structure is comparable to that of rLDPE, effectively suppressing the variability, with a $CV$ of only 8.8% (Figure 5C).

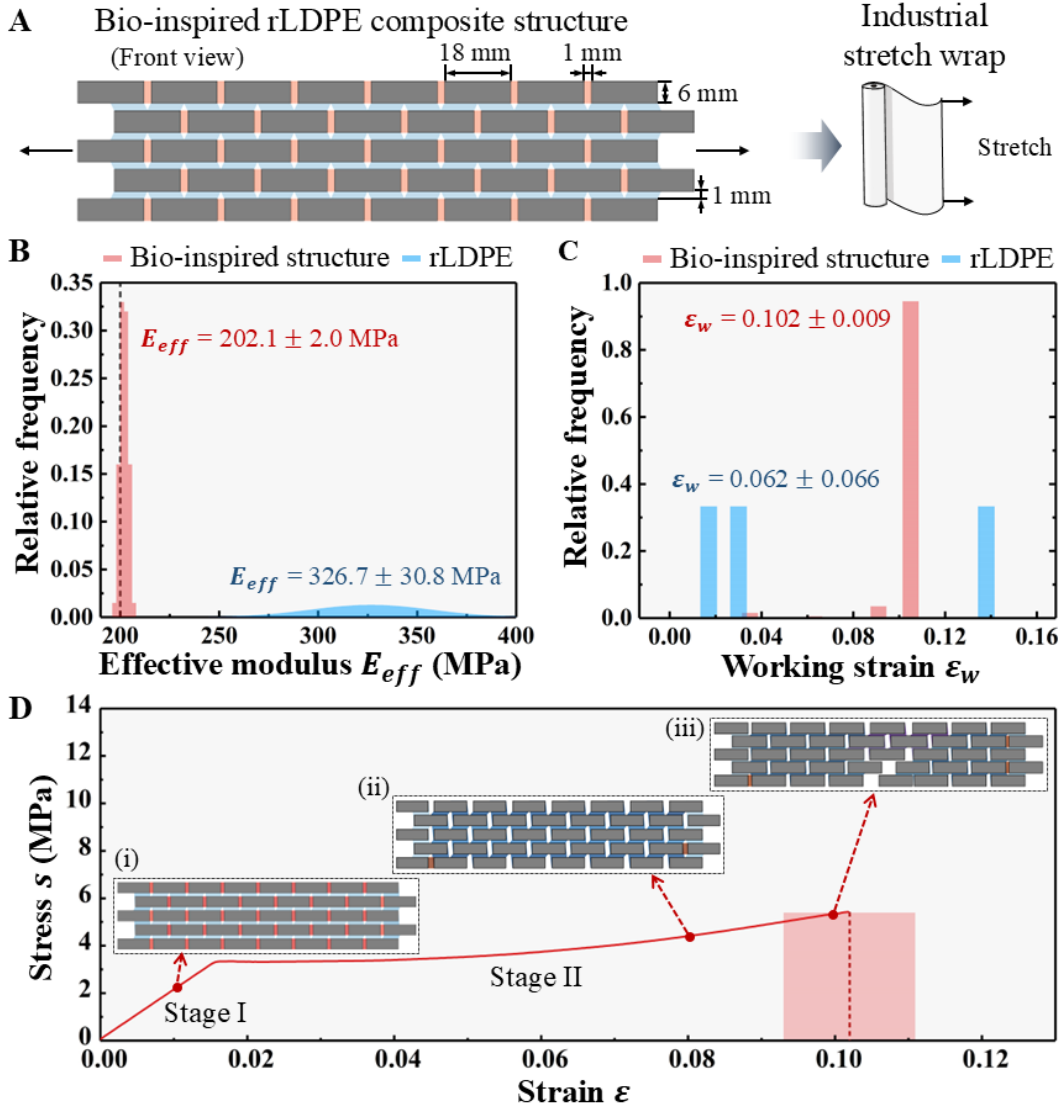

**Figure 5. Bio-inspired composite compared to rLDPE as industrial stretch wrap.** (**A**) The bio-inspired rLDPE composite structure, utilizing rLDPE as hard platelets, shows potential for use in industrial stretch wrap. (**B**) Frequency distribution in effective modulus $E_{eff}$ of bio-inspired structure (red) and rLDPE (blue) for 200 network simulations. The modulus of the virgin LDPE is 200MPa (dashed line). (**C**) Frequency distribution in working strain $\varepsilon_w$ of bio-inspired structure (red) and rLDPE (blue) for 200 network simulations. (**D**) Stress-strain curve of the bio-inspired structure. The effective modulus of the bio-inspired structure is 202.1MPa. The structure exhibits linear elastic behavior in Stage I. Inset (**i**) shows that the CT elements primarily carry the tensile load and undergo linear elastic deformation. In Stage II, the CT elements gradually fail, but the composite structure continues to undergo elastic deformation as the CS elements subsequently carry the load, as shown in inset (**ii**). When the CS elements experience significant

damage (shown in purple in inset (**iii**)), the entire structure exhibits irreversible damage. The working strain of bio-inspired structure ranges from 0.093 to 0.111.

The reduction in the effective modulus of the bio-inspired rLDPE composite structure effectively suppresses the adverse effects of material hardening caused by the recycling process. The effective modulus and working strain meet the requirements for the conformability of the industrial stretch wrap, making it suitable for adhering to products with complex shapes. The two-stage stress-strain curve of the bio-inspired structure is shown in Figure 5D. During Stage I, the bio-inspired structure behaves in a linear elastic manner, with the CT elements primarily bearing the load (as shown in inset (**i**)). As the strain increases, the CT elements reach their maximum stress, $\sigma_{max,CT}$, leading to gradual damage, which manifests as strain-softening during stage II. At this stage, the load is progressively transferred from the damaged CT elements to the CS elements, which remain in the elastic regime, allowing the composite to continue stretching (as shown in inset **ii**). Throughout this process, the failure of CT elements does not result in plastic deformation of the structure. When some CS elements enter the damage regime (shown as purple in inset (**iii**)), strain localization occurs, eventually leading to failure under a relatively small added load. Notably, the bio-inspired structure can be reused repeatedly after unloading, even after entering stage II, provided the CS elements are not severely damaged. Through this design, we reduced the modulus variability of rLDPE from 9.4% to less than 1%, while ensuring that the modulus of the bio-inspired structure meets the requirements of virgin LDPE. The effective working range of the bio-inspired structure spans from 0 to 0.102, comparable to that of rLDPE, while offering enhanced stability, improved conformability, and exhibiting nearly deterministic behavior. Altering the thickness of the stretch wrap would affect the load-bearing capacity of the structure but would not impact its working range or strain behavior. Therefore, the bio-inspired rLDPE composite meets the requirements for industrial stretch wrap.

## 4 Discussion and Conclusion

The bio-inspired composite structure design demonstrates significant potential in suppressing the variability inherent in recycled plastic materials through its brick-and-mortar architecture. Variability reduction is achieved through the complementary roles of CT and CS elements. CT elements, as primary load-bearing components, govern the structure's stiffness and elastic response, while CS elements, though secondary in stiffness contribution, play a crucial role in redistributing stress and accommodating deformation, particularly during progressive damage. Platelet geometry requires careful consideration; increasing platelet length slightly reduces stiffness due to the increased total length of the structure, while excessive platelet height yields diminishing returns because of cross-sectional area effects, rendering the aspect ratio effect unpredictable. Tailoring the mechanical properties of the hard and soft phases is equally critical. Soft, deterministic interfaces counteract the randomness of the hard phase by distributing deformation and delaying failure. Linear elastic hard platelets, while accommodating higher stresses with minimal contribution to overall deformation, could benefit from a large deformation model or a 2D framework to better capture their mechanical response under complex loading conditions such as those encountered in applications like stretch wraps. Overall, achieving desirable performance relies on a well-balanced interplay of the stiffness of CT elements, the compliance of CS interfaces, and the modulus of the hard platelets, ensuring robust and consistent results.

The mechanical properties of recyclates reported in the literature are highly inconsistent (see Figure 1A and outliers in Figure S1), with many studies featuring uncontrolled parameters that introduce significant uncertainty into the input data. This variability complicates accurate characterization and undermines the reliability of model predictions. To address these challenges, a dedicated experimental framework is essential to systematically evaluate recyclates' sensitivity to key parameters, including contamination, blends, recycling cycles, and processing conditions, while ensuring that experimental data is reported in a standardized, reproducible, and accessible format. Furthermore, coordinated guidelines are required to help researchers effectively navigate the extensive parameter space, enabling more consistent and reliable outcomes. The reliance on idealized interfacial properties further underscores the need for experimental validation to define traction-separation laws accurately. By integrating robust modeling approaches with a comprehensive understanding of recyclates' behavior and employing specialized high-throughput testing [46–53] the practical application of recyclates can be significantly advanced.

This study showcases the potential of bio-inspired composite structures to reduce variability in recycled plastics by integrating stochastic recyclate platelets as bricks with soft, deterministic polymeric interfaces as mortar. By carefully selecting network size, platelet dimensions and interfacial mechanical properties, the proposed design achieves improved stiffness and conformability while effectively suppressing material randomness. The model has been validated in two simple scenarios and applied in a case study about its potential use as an industrial wrapper, demonstrating comparable applicability with up to 89.5% reduction in stiffness variability and a 42% reduction in elongation at break variability.

Unlike other current efforts to reduce variability that heavily rely on chemistry-specific modifications [54,55], which can be challenging to scale and often pose significant environmental impacts [56], our bioinspired design approach is chemistry-agnostic. By leveraging universally applicable mechanics, it could offer a sustainable and versatile pathway to enhance the performance and reliability of recycled plastics.

## CRediT authorship contribution statement

**Dimitrios Georgiou:** Writing – review & editing, Writing – original draft, Visualization, Methodology, Investigation, Validation, Formal analysis, Software, Data curation, Conceptualization. **Danqi Sun:** Writing – review & editing, Writing – original draft, Visualization, Data Curation, Investigation. **Xing Liu:** Writing – review & editing, Writing – original draft, Methodology, Investigation, Validation, Formal analysis, Supervision. **Christos E. Athanasiou:** Writing – review & editing, Writing – original draft, Methodology, Investigation, Validation, Formal analysis, Conceptualization, Supervision, Funding acquisition.

## Declaration of Competing Interest

The authors declare that they have no known competing financial interests or personal relationships that could have appeared to influence the work reported in this paper.

## Acknowledgements

D.G. and C.E.A acknowledge support from the National Science Foundation (NSF) CAREER Award [CMMI-2338508].

## Data availability

Data will be made available on request.

## References

[1] Hannah Ritchie, Veronika Samborska and Max Roser (2023) - "Plastic Pollution" Published online at OurWorldinData.org. Retrieved from: "https://ourworldindata.org/plastic-pollution" [Online Resource], (n.d.).

[2] J.-G. Rosenboom, R. Langer, G. Traverso, Bioplastics for a circular economy, Nat Rev Mater 7 (2022) 117–137. https://doi.org/10.1038/s41578-021-00407-8.

[3] K.D. Nixon, Z.O.G. Schyns, Y. Luo, M.G. Ierapetritou, D.G. Vlachos, L.T.J. Korley, T.H. Epps, III, Analyses of circular solutions for advanced plastics waste recycling, Nature Chemical Engineering 1 (2024) 615–626. https://doi.org/10.1038/s44286-024-00121-6.

[4] Z.O.G. Schyns, M.P. Shaver, Mechanical Recycling of Packaging Plastics: A Review, Macromol Rapid Commun 42 (2021). https://doi.org/10.1002/marc.202000415.

[5] C.E. Athanasiou, X. Liu, H. Gao, A Perspective on Democratizing Mechanical Testing: Harnessing Artificial Intelligence to Advance Sustainable Material Adoption and Decentralized Manufacturing, J Appl Mech 91 (2024). https://doi.org/10.1115/1.4066085.

[6] F.P. La Mantia, M. Vinci, Recycling poly(ethyleneterephthalate), Polym Degrad Stab 45 (1994) 121–125. https://doi.org/10.1016/0141-3910(94)90187-2.

[7]  M.K. Eriksen, J.D. Christiansen, A.E. Daugaard, T.F. Astrup, Closing the loop for PET, PE and PP waste from households: Influence of material properties and product design for plastic recycling, Waste Management 96 (2019) 75–85. https://doi.org/10.1016/j.wasman.2019.07.005.

[8]  V.S. Cecon, G.W. Curtzwiler, K.L. Vorst, Evaluation of mixed #3–7 plastic waste from material recovery facilities (MRFs) in the United States, Waste Management 171 (2023) 313–323. https://doi.org/10.1016/j.wasman.2023.09.002.

[9]  H.D. Espinosa, J.E. Rim, F. Barthelat, M.J. Buehler, Merger of structure and material in nacre and bone – Perspectives on de novo biomimetic materials, Prog Mater Sci 54 (2009) 1059–1100. https://doi.org/10.1016/j.pmatsci.2009.05.001.

[10] S.W. Cranford, A. Tarakanova, N.M. Pugno, M.J. Buehler, Nonlinear material behaviour of spider silk yields robust webs, Nature 482 (2012) 72–76. https://doi.org/10.1038/nature10739.

[11] C. Fox, K. Chen, M. Antonini, T. Magrini, C. Daraio, Extracting Geometry and Topology of Orange Pericarps for the Design of Bioinspired Energy Absorbing Materials, Advanced Materials (2024). https://doi.org/10.1002/adma.202405567.

[12] A. Chen, U. Ezimora, S. Lee, J.-H. Lee, G.X. Gu, Sea sponge-inspired designs enhance mechanical properties of tubular lattices, Int J Mech Sci 285 (2025) 109815. https://doi.org/10.1016/j.ijmecsci.2024.109815.

[13] H. Gao, B. Ji, I.L. Jäger, E. Arzt, P. Fratzl, Materials become insensitive to flaws at nanoscale: Lessons from nature, Proceedings of the National Academy of Sciences 100 (2003) 5597–5600. https://doi.org/10.1073/pnas.0631609100.

[14] A. Jackson, J. Vincent, R. Turner, The mechanical design of nacre, Proc R Soc Lond B Biol Sci 234 (1988) 415–440. https://doi.org/10.1098/rspb.1988.0056.

[15] A.G. Evans, Z. Suo, R.Z. Wang, I.A. Aksay, M.Y. He, J.W. Hutchinson, Model for the robust mechanical behavior of nacre, J Mater Res 16 (2001) 2475–2484. https://doi.org/10.1557/JMR.2001.0339.

[16] R.Z. Wang, Z. Suo, A.G. Evans, N. Yao, I.A. Aksay, Deformation mechanisms in nacre, J Mater Res 16 (2001) 2485–2493. https://doi.org/10.1557/JMR.2001.0340.

[17] J. Sun, B. Bhushan, Hierarchical structure and mechanical properties of nacre: A review, RSC Adv 2 (2012) 7617–7632. https://doi.org/10.1039/c2ra20218b.

[18] Y. Yan, Z.L. Zhao, X.Q. Feng, H. Gao, Nacre's brick–mortar structure suppresses the adverse effect of microstructural randomness, J Mech Phys Solids 159 (2022). https://doi.org/10.1016/j.jmps.2021.104769.

[19] F. Barthelat, C.-M. Li, C. Comi, H.D. Espinosa, Mechanical properties of nacre constituents and their impact on mechanical performance, J Mater Res 21 (2006) 1977–1986. https://doi.org/10.1557/jmr.2006.0239.

[20] C. Pattanakul, S. Selke, C. Lai, J. Miltz, Properties of recycled high density polyethylene from milk bottles, J Appl Polym Sci 43 (1991) 2147–2150. https://doi.org/10.1002/app.1991.070431122.

[21] A. Boldizar, A. Jansson, T. Gevert, K. Möller, Simulated recycling of post-consumer high density polyethylene material, Polym Degrad Stab 68 (2000) 317–319. https://doi.org/10.1016/S0141-3910(00)00012-4.

[22] H.M. da Costa, V.D. Ramos, M.G. de Oliveira, Degradation of polypropylene (PP) during multiple extrusions: Thermal analysis, mechanical properties and analysis of variance, Polym Test 26 (2007) 676–684. https://doi.org/10.1016/j.polymertesting.2007.04.003.

[23] P. Brachet, L.T. Høydal, E.L. Hinrichsen, F. Melum, Modification of mechanical properties of recycled polypropylene from post-consumer containers, Waste Management 28 (2008) 2456–2464. https://doi.org/10.1016/j.wasman.2007.10.021.

[24] A.A. Mendes, A.M. Cunha, C.A. Bernardo, Study of the degradation mechanisms of polyethylene during reprocessing, Polym Degrad Stab 96 (2011) 1125–1133. https://doi.org/10.1016/j.polymdegradstab.2011.02.015.

[25] H. Jin, J. Gonzalez-Gutierrez, P. Oblak, B. Zupančič, I. Emri, The effect of extensive mechanical recycling on the properties of low density polyethylene, Polym Degrad Stab 97 (2012) 2262–2272. https://doi.org/10.1016/j.polymdegradstab.2012.07.039.

[26] P. Oblak, J. Gonzalez-Gutierrez, B. Zupančič, A. Aulova, I. Emri, Processability and mechanical properties of extensively recycled high density polyethylene, Polym Degrad Stab 114 (2015) 133–145. https://doi.org/10.1016/j.polymdegradstab.2015.01.012.

[27] J. Aurrekoetxea, M.A. Sarrionandia, I. Urrutibeascoa, M.Ll. Maspoch, Effects of recycling on the microstructure and the mechanical properties of isotactic polypropylene, J Mater Sci 36 (2001) 2607–2613. https://doi.org/10.1023/A:1017983907260.

[28] B. Ji, H. Gao, Mechanical properties of nanostructure of biological materials, J Mech Phys Solids 52 (2004) 1963–1990. https://doi.org/10.1016/j.jmps.2004.03.006.

[29] W. Luo, Z.P. Bažant, Fishnet model for failure probability tail of nacre-like imbricated lamellar materials, Proc Natl Acad Sci U S A 114 (2017) 12900–12905. https://doi.org/10.1073/pnas.1714103114.

[30] L. Léger, C. Creton, Adhesion mechanisms at soft polymer interfaces, Philosophical Transactions of the Royal Society A: Mathematical, Physical and Engineering Sciences 366 (2008) 1425–1442. https://doi.org/10.1098/rsta.2007.2166.

[31] C. Verdier, G. Ravilly, Peeling of polydimethylsiloxane adhesives: The case of adhesive failure, J Polym Sci B Polym Phys 45 (2007) 2113–2122. https://doi.org/10.1002/polb.21045.

[32] T. Zhang, H. Yuk, S. Lin, G.A. Parada, X. Zhao, Tough and tunable adhesion of hydrogels: experiments and models, Acta Mechanica Sinica 33 (2017) 543–554. https://doi.org/10.1007/s10409-017-0661-z.

[33] A. Khayer Dastjerdi, E. Tan, F. Barthelat, Direct Measurement of the Cohesive Law of Adhesives Using a Rigid Double Cantilever Beam Technique, Exp Mech 53 (2013) 1763–1772. https://doi.org/10.1007/s11340-013-9755-0.

[34] Chaz Miller (2010) - "Plastic Film 6025" Published online at waste360.com. Retrieved from: "https://www.waste360.com/waste-recycling/plastic-film-6025" [Online Resource], (n.d.).

[35] Fortune Business Insights, Stretch Wrap Market Size, Share & Industry Analysis, By Type (Cast Wrap and Blown Wrap), By Material (Polyethylene (PE), [Low Density Polyethylene (LDPE) and High Density Polyethylene (HDPE)], Polyvinyl Chloride (PVC), Polypropylene (PP), Biaxially Oriented Polypropylene (BOPP), and Others), By End User (Shipping & Logistics, Food & Beverage, Agriculture, Building & Construction, Chemical, Household, and Others), and Regional Forecast, 2024-2032., 2025.

[36] V.P. Ranjan, S. Goel, Degradation of Low-Density Polyethylene Film Exposed to UV Radiation in Four Environments, J Hazard Toxic Radioact Waste 23 (2019). https://doi.org/10.1061/(ASCE)HZ.2153-5515.0000453.

[37] G. Ayoub, A.K. Rodriguez, B. Mansoor, X. Colin, Modeling the visco-hyperelastic–viscoplastic behavior of photodegraded semi-crystalline low-density polyethylene films, Int J Solids Struct 204–205 (2020) 187–198. https://doi.org/10.1016/j.ijsolstr.2020.08.025.

[38] M. Rujnić Havstad, I. Tucman, Z. Katančić, A. Pilipović, Influence of Ageing on Optical, Mechanical, and Thermal Properties of Agricultural Films, Polymers (Basel) 15 (2023) 3638. https://doi.org/10.3390/polym15173638.

[39] E.U. Thoden van Velzen, S. Chu, F. Alvarado Chacon, M.T. Brouwer, K. Molenveld, The impact of impurities on the mechanical properties of recycled polyethylene, Packaging Technology and Science 34 (2021) 219–228. https://doi.org/10.1002/pts.2551.

[40] S.R. Mellott, A. Fatemi, Fatigue behavior and modeling of thermoplastics including temperature and mean stress effects, Polym Eng Sci 54 (2014) 725–738. https://doi.org/10.1002/pen.23591.

[41] C. Meran, O. Ozturk, M. Yuksel, Examination of the possibility of recycling and utilizing recycled polyethylene and polypropylene, Mater Des 29 (2008) 701–705. https://doi.org/10.1016/j.matdes.2007.02.007.

[42] A.G. Pedroso, D.S. Rosa, Mechanical, thermal and morphological characterization of recycled LDPE/corn starch blends, Carbohydr Polym 59 (2005) 1–9. https://doi.org/10.1016/j.carbpol.2004.08.018.

[43] S. Saikrishnan, D. Jubinville, C. Tzoganakis, T.H. Mekonnen, Thermo-mechanical degradation of polypropylene (PP) and low-density polyethylene (LDPE) blends exposed to simulated recycling, Polym Degrad Stab 182 (2020) 109390. https://doi.org/10.1016/j.polymdegradstab.2020.109390.

[44] G.-F. Shan, W. Yang, B.-H. Xie, M.-B. Yang, Mechanical Properties and Morphology of LDPE/PP Blends, Journal of Macromolecular Science, Part B 46 (2007) 963–974. https://doi.org/10.1080/00222340701457253.

[45] C.L. Choy, W.P. Leung, H.C. Ng, Mechanical relaxations and moduli of isotropic and oriented linear low-density polyethylene, J Appl Polym Sci 32 (1986) 5883–5901. https://doi.org/10.1002/app.1986.070320719.

[46]     C.-E. Athanasiou, Y. Bellouard, A Monolithic Micro-Tensile Tester for Investigating Silicon Dioxide Polymorph Micromechanics, Fabricated and Operated Using a Femtosecond Laser, Micromachines (Basel) 6 (2015) 1365–1386. https://doi.org/10.3390/mi6091365.

[47]     C.-E. Athanasiou, M.-O. Hongler, Y. Bellouard, Unraveling Brittle-Fracture Statistics from Intermittent Patterns Formed During Femtosecond Laser Exposure, Phys Rev Appl 8 (2017) 054013. https://doi.org/10.1103/PhysRevApplied.8.054013.

[48]     C.E. Athanasiou, X. Liu, B. Zhang, T. Cai, C. Ramirez, N.P. Padture, J. Lou, B.W. Sheldon, H. Gao, Integrated simulation, machine learning, and experimental approach to characterizing fracture instability in indentation pillar-splitting of materials, J Mech Phys Solids 170 (2023) 105092. https://doi.org/10.1016/j.jmps.2022.105092.

[49]     C.E. Athanasiou, C.D. Fincher, C. Gilgenbach, H. Gao, W.C. Carter, Y.-M. Chiang, B.W. Sheldon, Operando measurements of dendrite-induced stresses in ceramic electrolytes using photoelasticity, Matter 7 (2024) 95–106. https://doi.org/10.1016/j.matt.2023.10.014.

[50]     S.I. Nazir, C.E. Athanasiou, Y. Bellouard, On the behavior of uniaxial static stress loaded micro-scale fused silica beams at room temperature, Journal of Non-Crystalline Solids: X 14 (2022) 100083. https://doi.org/10.1016/j.nocx.2022.100083.

[51]     X. Liu, C.E. Athanasiou, N.P. Padture, B.W. Sheldon, H. Gao, A machine learning approach to fracture mechanics problems, Acta Mater 190 (2020) 105–112. https://doi.org/10.1016/j.actamat.2020.03.016.

[52]     X. Liu, C.E. Athanasiou, N.P. Padture, B.W. Sheldon, H. Gao, Knowledge extraction and transfer in data-driven fracture mechanics, Proceedings of the National Academy of Sciences 118 (2021). https://doi.org/10.1073/pnas.2104765118.

[53]     A.D. Patel, Z.O.G. Schyns, T.W. Franklin, M.P. Shaver, Defining quality by quantifying degradation in the mechanical recycling of polyethylene, Nat Commun 15 (2024) 8733. https://doi.org/10.1038/s41467-024-52856-8.

[54]     T. Vialon, H. Sun, G.J.M. Formon, P. Galanopoulo, C. Guibert, F. Averseng, M.-N. Rager, A. Percot, Y. Guillaneuf, N.J. Van Zee, R. Nicolaÿ, Upcycling Polyolefin Blends into High-Performance Materials by Exploiting Azidotriazine Chemistry Using Reactive Extrusion, J Am Chem Soc 146 (2024) 2673–2684. https://doi.org/10.1021/jacs.3c12303.

[55]     P.M. Stathatou, L. Corbin, J.C. Meredith, A. Garmulewicz, Biomaterials and Regenerative Agriculture: A Methodological Framework to Enable Circular Transitions, Sustainability 15 (2023) 14306. https://doi.org/10.3390/su151914306.

[56]     L. Yang, H. Li, Y. Zhang, N. Jiao, Environmental risk assessment of triazine herbicides in the Bohai Sea and the Yellow Sea and their toxicity to phytoplankton at environmental concentrations, Environ Int 133 (2019) 105175. https://doi.org/10.1016/j.envint.2019.105175.

# Suppressing Mechanical Property Variability in Recycled Plastics *via* Bio-inspired Design

Dimitrios Georgiou[1], Danqi Sun[1], Xing Liu[2], Christos E Athanasiou[1,*]

1. Daniel Guggenheim School of Aerospace Engineering, Georgia Institute of Technology, Atlanta, GA 30332, USA

2. Department of Mechanical and Industrial Engineering, New Jersey Institute of Technology, Newark, NJ 07102, USA

*Corresponding author: Christos E Athanasiou (athanasiou@gatech.edu)

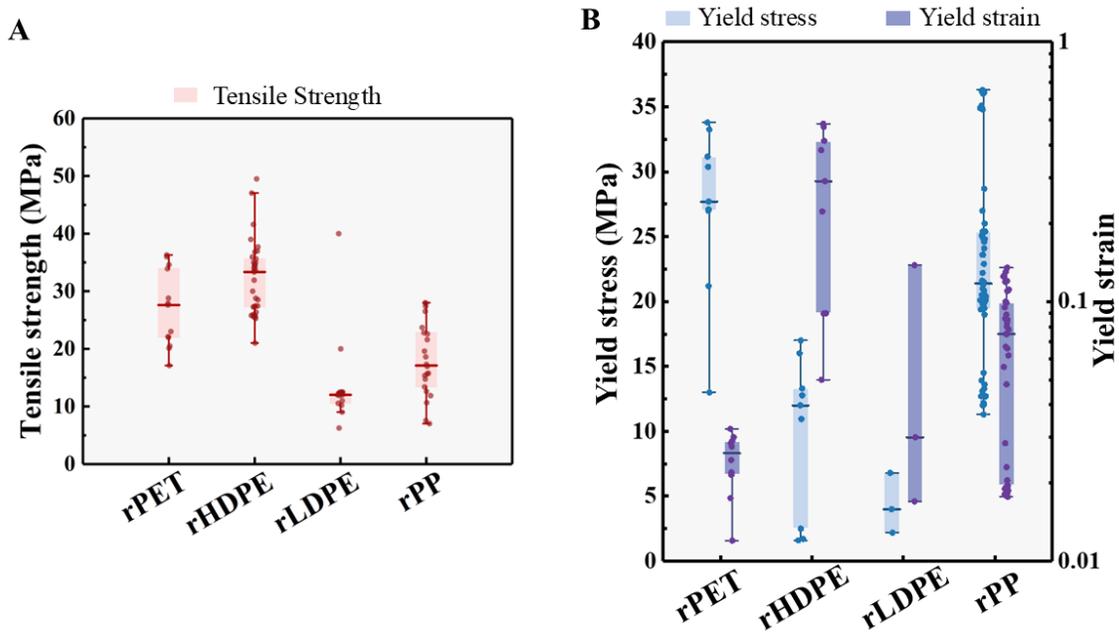

**Figure S1. Literature Data on the mechanical properties of Recycled Plastics. (A)** *Tensile Strength:* Compilation of tensile strength values from various studies, illustrating the broad variability in recycled plastics. Several outliers are noticeable, deviating significantly from the main trend, likely due to differences in material composition, contamination levels, and processing conditions. The lack of standardization across studies results in inconsistencies that complicate direct comparisons. **(B)** *Yield Point:* Yield stress and strain data displaying notable scatter across different recycling conditions. The rLDPE category contains scarce data, limiting statistical reliability, while others exhibit significant variability, emphasizing the influence of material blends, number of recycling cycles, and testing conditions. The variability observed also reinforces the need for systematic experimental frameworks to better quantify and predict the mechanical behavior of recyclates. All data were drawn from [1–11]

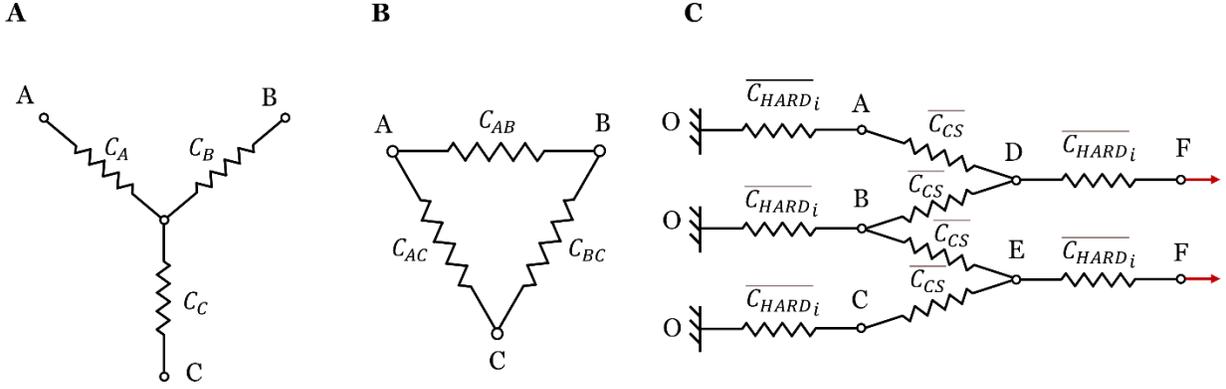

**Figure S2. Analysis of the Spring Connectivity of the Alternating Column case. (A)** Wye ($Y$) spring connectivity **(B)** Delta ($\Delta$) spring connectivity **(C)** Connectivity of a {1,5} alternating column network. The alternating column configuration features a mixed series-parallel connectivity that can be analyzed using these transformations. Considering the compliance $C$ of an element, it is possible to transform a $Y$ configuration into an equivalent $\Delta$ one and vice versa

<u>$Y$ to $\Delta$ Transformation</u>

$$C_{AB} = \frac{C_A C_B + C_B C_C + C_C C_A}{C_C}$$

$$C_{BC} = \frac{C_A C_B + C_B C_C + C_C C_A}{C_A}$$

$$C_{CA} = \frac{C_A C_B + C_B C_C + C_C C_A}{C_B}$$

<u>$\Delta$ to $Y$ Transformation</u>

$$C_A = \frac{C_{AB} C_{CA}}{C_{AB} + C_{BC} + C_{CA}}$$

$$C_B = \frac{C_{AB} C_{BC}}{C_{AB} + C_{BC} + C_{CA}}$$

$$C_C = \frac{C_{BC} C_{CA}}{C_{AB} + C_{BC} + C_{CA}}$$

All fixed nodes are reduced to a single node (O), while all displacement nodes collapse into a separate node (F). By systematically simplifying the network through series and parallel reductions and applying $Y-to-\Delta$ and $\Delta-to-Y$ transformations where necessary, the entangled connectivity is resolved. This process ultimately yields an equivalent stiffness value, as expressed in Eq. (15).

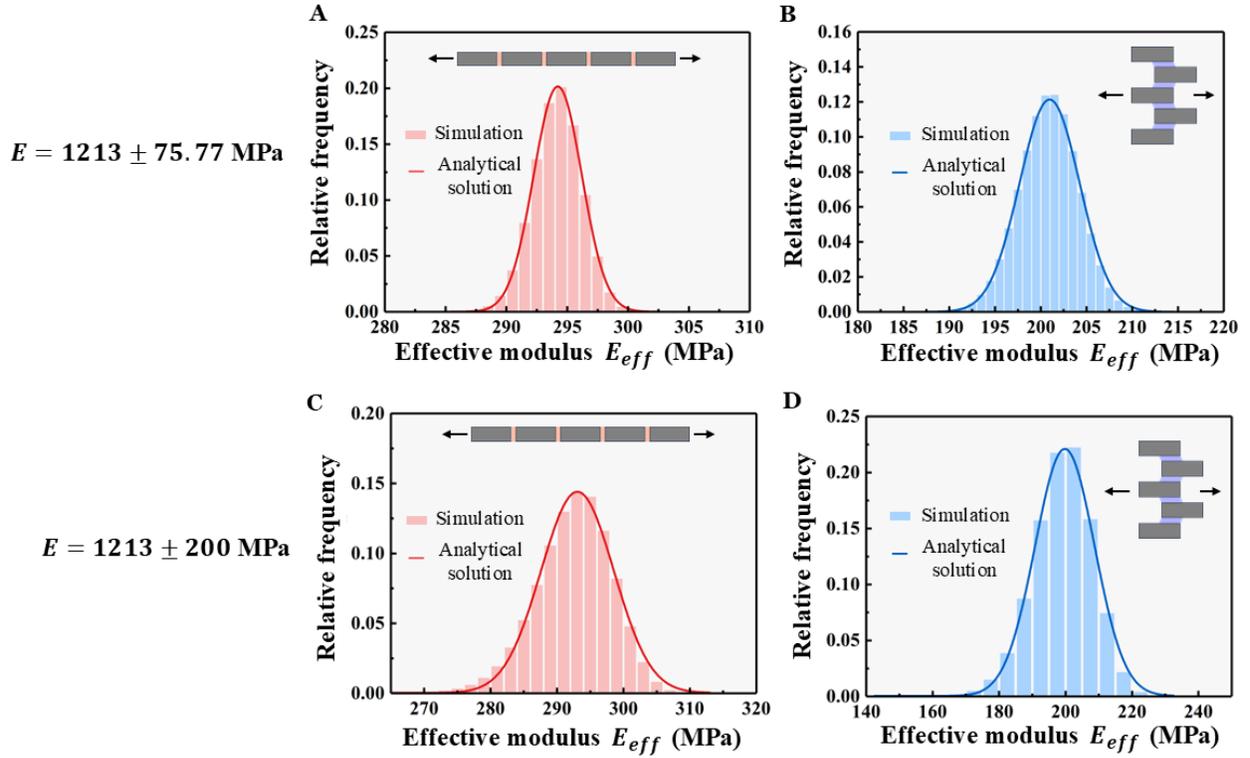

**Figure S3. Effect of the magnitude of variability on the network's distribution (A)** presents the effective modulus distribution for the serial chain configuration under low variability, where stiffness is primarily governed by the most compliant constituent. **(B)** shows the corresponding distribution for the alternating column configuration, where stiffness arises from the combined contributions of hard elements and CS interfaces (see Figure S2 and Eq. (15)). **(C)** and **(D)** depict the same configurations under higher input variability (same plots as Figure 3). In (C), the serial chain model exhibits a pronounced left skew, as weaker platelets disproportionately reduce stiffness, while stronger platelets contribute minimally. (D) shows a similar skew for the alternating column configuration, albeit less pronounced due to its column-wise connectivity. Comparing (A, B) to (C, D), we see that increased variability results in a left tail, highlighting the heightened sensitivity of the network to the compliant constituents.

The skewness values indicate that both configurations exhibit left-skewed distributions, meaning that weaker elements disproportionately affect the overall stiffness, a trend more pronounced in the serial chain model. This asymmetry is further reflected in the excess kurtosis, which suggests that the serial chain configuration has a sharper peak and heavier tails compared to the alternating column case, indicating a higher likelihood of extreme values. The goodness-of-fit metrics ($R^2$) for the normal and Weibull distributions reveal that the effective modulus follows a near-normal distribution, with slightly better fits for the normal distribution in both configurations, validating the statistical approach used in this analysis

**Table S1. Comparison of distribution properties between the material, the semi-analytical solution and the developed model for the validation cases of the original manuscript**

| *Metric* | *Material* | *Serial Chain* | | *Alternating Column* | |
|---|---|---|---|---|---|
| | | *Analytical* | *Model* | *Analytical* | *Model* |
| Mean $\mu$ | 1213.84 | 292.49 | 292.48 | 199.15 | 199.13 |
| Standard Deviation $\sigma$ | 200.00 | 5.6714 | 5.6726 | 9.0418 | 8.7370 |
| $CV$ | 16.49% | 1.94% | 1.94% | 4.54% | 4.39% |
| Skewness | 0 | -0.4496 | -0.4423 | -0.2980 | -0.3073 |
| Excess Kurtosis | 0 | 0.5075 | 0.4895 | 0.2126 | 0.2379 |
| $R^2$ value (Normal) | 1.00 | 0.9875 | 0.9875 | 0.9934 | 0.9935 |
| $R^2$ value (Weibull) | N/A | 0.9789 | 0.9767 | 0.9721 | 0.9730 |

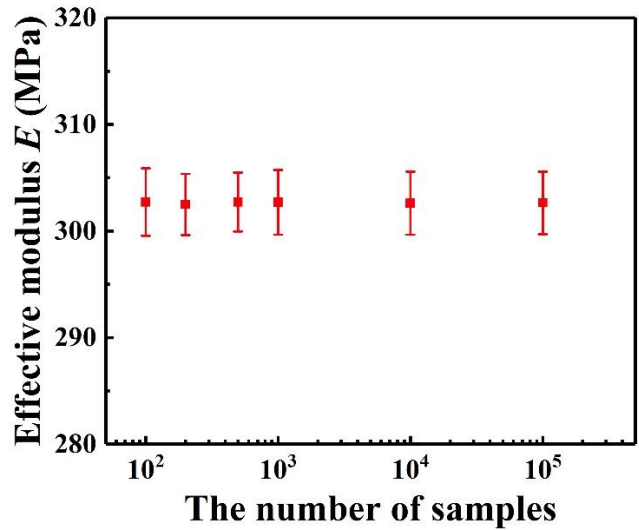

**Figure S4. Convergence analysis.** The plot illustrates the convergence of the effective modulus $E_{eff}$ and its variability as a function of the number of Monte Carlo simulations for a network of size {10,5}. Both the mean effective modulus (red squares) and the standard deviation (error bars) exhibit minor fluctuations, reflecting statistical confidence and showcasing that $10^2$ samples provide a sufficiently converged estimate of $E_{eff}$ in both magnitude and dispersion.

[1]   C. Pattanakul, S. Selke, C. Lai, J. Miltz, Properties of recycled high density polyethylene from milk bottles, J Appl Polym Sci 43 (1991) 2147–2150. https://doi.org/10.1002/app.1991.070431122.

[2]   F.P. La Mantia, M. Vinci, Recycling poly(ethyleneterephthalate), Polym Degrad Stab 45 (1994) 121–125. https://doi.org/10.1016/0141-3910(94)90187-2.

[3]   A. Boldizar, A. Jansson, T. Gevert, K. Möller, Simulated recycling of post-consumer high density polyethylene material, Polym Degrad Stab 68 (2000) 317–319. https://doi.org/10.1016/S0141-3910(00)00012-4.

[4]   H.M. da Costa, V.D. Ramos, M.G. de Oliveira, Degradation of polypropylene (PP) during multiple extrusions: Thermal analysis, mechanical properties and analysis of variance, Polym Test 26 (2007) 676–684. https://doi.org/10.1016/j.polymertesting.2007.04.003.

[5]   P. Brachet, L.T. Høydal, E.L. Hinrichsen, F. Melum, Modification of mechanical properties of recycled polypropylene from post-consumer containers, Waste Management 28 (2008) 2456–2464. https://doi.org/10.1016/j.wasman.2007.10.021.

[6]   A.A. Mendes, A.M. Cunha, C.A. Bernardo, Study of the degradation mechanisms of polyethylene during reprocessing, Polym Degrad Stab 96 (2011) 1125–1133. https://doi.org/10.1016/j.polymdegradstab.2011.02.015.

[7]   H. Jin, J. Gonzalez-Gutierrez, P. Oblak, B. Zupančič, I. Emri, The effect of extensive mechanical recycling on the properties of low density polyethylene, Polym Degrad Stab 97 (2012) 2262–2272. https://doi.org/10.1016/j.polymdegradstab.2012.07.039.

[8]   P. Oblak, J. Gonzalez-Gutierrez, B. Zupančič, A. Aulova, I. Emri, Processability and mechanical properties of extensively recycled high density polyethylene, Polym Degrad Stab 114 (2015) 133–145. https://doi.org/10.1016/j.polymdegradstab.2015.01.012.

[9]   J. Aurrekoetxea, M.A. Sarrionandia, I. Urrutibeascoa, M.Ll. Maspoch, Effects of recycling on the microstructure and the mechanical properties of isotactic polypropylene, J Mater Sci 36 (2001) 2607–2613. https://doi.org/10.1023/A:1017983907260.

[10]  M.K. Eriksen, J.D. Christiansen, A.E. Daugaard, T.F. Astrup, Closing the loop for PET, PE and PP waste from households: Influence of material properties and product design for plastic recycling, Waste Management 96 (2019) 75–85. https://doi.org/10.1016/j.wasman.2019.07.005.

[11]  V.S. Cecon, G.W. Curtzwiler, K.L. Vorst, Evaluation of mixed #3–7 plastic waste from material recovery facilities (MRFs) in the United States, Waste Management 171 (2023) 313–323. https://doi.org/10.1016/j.wasman.2023.09.002.